\documentclass{rstransa}
\jname{rsta}
\Journal{Phil.\ Trans.\ R.\ Soc.}

\begin{document}

\title{Stochastic parametric excitation of convective heat transfer}

\author{
Evelina V.\ Permyakova$^{1,2}$ and Denis S.\ Goldobin$^{1,2}$}

\address{$^{1}$Institute of Continuous Media Mechanics, UB RAS, Academician Korolev Street 1, 614013 Perm, Russia\\
$^{2}$Department of Theoretical Physics, Perm State University, Bukirev Street 15, 614990 Perm, Russia}

\subject{Fluid Dynamics, Statistical Physics, Nonlinear Dynamics}

\corres{Denis S. Goldobin\\
\email{Denis.Goldobin@gmail.com}}

\begin{abstract}
We study the parametric excitation of the free thermal convection in a horizontal layer and a rectangular cell by random vertical vibrations. The mathematical formulation we use allows one to explore the cases of heating from below and above and the low-gravity conditions. The excitation threshold of the second moments of the current velocity and the temperature perturbations is derived. The heat flux through the system quantified by the Nusselt number is reported to be related to the second moment of temperature perturbations; therefore, the threshold of the stochastic excitation of second moments gives the threshold for the excitation of the convective heat transfer. Comparison of the stochastic parametric excitation to the effect of high-frequency periodic modulation reveals dramatic dissimilarity between the two.
\end{abstract}

\begin{fmtext}
\section{Introduction}
The topic of random parametric excitation (or {\it stochastic stability}) in fluid mechanics is not as broadly addressed as its deterministic counterpart of a periodic modulation (e.g., see~\cite{Ibrahim-2015}). The research interest to the former was triggered by the beginning of the spaceflights era and further boosted by the low-gravity experiments, where an extreme sensitivity to the fluctuations of the residual gravity acceleration was encountered~\cite{Alexander-1990}.

Primary attention here was devoted to the parametric sloshing by random vibrations (see~\cite{Mitchell-1968}, review~\cite{Ibrahim-2015} and references therein). The important peculiarity of the stochastic stability subject is the fact that, technically, one finds different excitation conditions for the moments of different order~\cite{Khasminskii-2012}. In fluid dynamics this peculiarity was
\end{fmtext}

\maketitle

\noindent
well recognized already in pioneering works; for instance, Mitchell~\cite{Mitchell-1968} clearly distinguished the stability of moments and the almost sure stability. Physically, for the problem of sloshing, different instability thresholds for different quantifiers give rise to four different regimes~\cite{Ibrahim-Heinrich-1988,Ibrahim-2015}; in order of appearance as the noise intensity increases, (a)~{\it zero motion of the free liquid surface}, (b)~{\it on--off intermittent motion} of the free liquid surface, characterized by very small motion whenever it occurs, (c)~{\it partially developed random sloshing}, when significant free surface motion occurs for a certain time-interval and then ceases for another interval, (d)~{\it fully developed sloshing} characterized by continuous random liquid motion.

In the literature, theoretical studies of the random parametric sloshing were mostly focused on the case of a narrow-band residual gravity acceleration
 $g(t)=a\cos\Omega t+\xi_1(t)$,
as adopted in the earliest works~\cite{Mitchell-1968}, or
 $g(t)=\xi_1(t)\cos\Omega t+\xi_2(t)\sin\Omega t$ in~\cite{Zhang-etal-1993,Ibrahim-2015}, where $\xi_1(t)$ and $\xi_2(t)$ are independent Gaussian noises.
Forced thermal convection under the narrow-band $g$-jitter was studied in~\cite{Thomson-etal-1995}; this case is mathematically far from the subject of stochastic stability, where trivial solutions remain possible.

For the excitation of the counterpart of the Faraday ripples~\cite{Faraday-1831}---random parametric sloshing---the interest to the case of a narrow-band noise is natural. For instance, one can deal with the excitation of Faraday ripples, which is a parametric resonance phenomenon~(e.g., see~\cite{Birikh-etal-2001}), distorted by a practically irreducible noise under microgravity conditions. For the case of the effect of random vibrations of the thermal convection system, there is no essential need for the periodic component of modulation. Even though some researchers consider here a narrow-band noise $g(t)=\xi_1(t)\cos\Omega t+\xi_2(t)\sin\Omega t$~\cite{Thomson-etal-1995}, the Fourier-spectra of the residual gravity acceleration in spacelabs in the range from 0 to 200~Hz~\cite{Knabe-Eilers-1982} are rather more resemblant of the white noise than of a superposition of a few narrow-band noises. In~\cite{Knabe-Eilers-1982}, the range from 0 to 10~Hz, where one can think of a narrow-band noise, is not only a small fraction of the whole analysed range 0--200~Hz, but also was seriously affected by the Hanning normalized function with a duration of 2.56~sec. Therefore, we consider the case of the white-noise approximation for the irregular component of vibrations to be worthy of study.

In this paper we consider the excitation of natural thermal convection in an infinite horizontal layer or a rectangular cavity by vertical random vibrations. Our analysis is not restricted to the case of microgravity conditions. Our ultimate results in this paper are the conditions for the excitation of second moments. We argue that this is a suitable characterization since the convective heat transfer through the system is quadratic with respect to the stream function amplitude and temperature perturbation. In section~\ref{sec:Nu}, the specific relation between the Nusselt number and the second moments is reported.

The paper is organized as follows. In section~\ref{sec:thc}, we formulate the specific physical problem we consider and derive dimensionless stochastic ODEs for linear perturbations. In section~\ref{sec:spe}, we recall the derivation of the moment equations for stochastic linear ODEs. In section~\ref{sec:spethc}, the moment equations are derived and analysed specifically for the thermal convection; the conditions for the excitation of the convective heat transfer are obtained. In section~\ref{sec:Nu}, we derive the relation between the Nusselt number and second moments calculated in section~\ref{sec:spethc}. In section~\ref{sec:periodic}, the mechanism of the stochastic parametric excitation is compared to that of the high-frequency periodic modulation. Section~\ref{sec:concl} finalizes the paper with conclusion.

\section{Thermal convection instability under random modulation of gravity acceleration}\label{sec:thc}
In this section we recall the mathematical formulation of the Rayleigh--B\'enard convection problem~\cite{Rayleigh-1916,Benard-1900,Gershuni-Zhukhovitskii-1976} and introduce random modulation of the gravity acceleration in the system.
We consider a horizontal fluid layer of thickness $h$. The linear stability of this system can be treated within the $(x,z)$-geometry with the $z$-axis directed vertically upwards; flows are assumed to be homogeneous along the $y$-axis. With the Boussinesq approximation one adopts a linear dependence of the fluid density $\rho$ on temperature $T$, $\rho(T)=\rho_0(1-\beta T)$, where $\beta$ is the thermal expansion coefficient, in the buoyancy force and neglects the density variation in other terms. In this case, the momentum conservation, mass conservation, and heat transfer equations read
\begin{align}
\rho_0\left(\frac{\partial\mathbf{v}}{\partial t} +(\mathbf{v}\cdot\nabla)\mathbf{v}\right)
&=-\nabla p+\rho_0\nu\triangle\mathbf{v}-\rho_0\beta T\mathbf{g}\,,
\label{eq:tc01}
\\
\nabla\cdot\mathbf{v}&=0\,,
\label{eq:tc02}
\\
\frac{\partial T}{\partial t} +(\mathbf{v}\cdot\nabla)T
&=\chi\triangle T\,,
\label{eq:tc03}
\end{align}
where the fluid flow velocity $\mathbf{v}=\{v^{(x)},0,v^{(z)}\}$, Laplace operator $\triangle=\frac{\partial^2}{\partial x^2}+\frac{\partial^2}{\partial z^2}$, $p$: pressure, $\nu$: kinematic viscosity, $\mathbf{g}$: the gravity acceleration which can be modulated in time, $\chi$: heat diffusivity.

For a linear stability of the heat conducting state, $\mathbf{v}$ is small and $T=-sAz+\theta$, where $A$ is the absolute value of the temperature gradient in the conducting state, $s=-1$ for heating from above and $s=+1$ for heating from below, temperature perturbation $\theta$ is small. After rescaling $(x,z)\to(hx,hz)$, $t\to(h^2/\nu)t$, $v\to(\chi/h)v$, $\theta\to Ah\theta$, and $p\to(\rho_0\nu\chi/h^2)p$, the linearization of equations (\ref{eq:tc01}) and (\ref{eq:tc03}) can be recast in the dimensionless form:
\begin{align}
\frac{\partial\mathbf{v}}{\partial t}
&=-\nabla p+\triangle\mathbf{v}+\mathrm{Ra}\theta\mathbf{e}_z\,,
\label{eq:tc04}
\\
\frac{\partial\theta}{\partial t}
&=\frac{1}{\mathrm{Pr}}\left(\triangle\theta+sv^{(z)}\right),
\label{eq:tc05}
\end{align}
where $\mathrm{Ra}=g\beta Ah^4/(\nu\chi)$ is the Rayleigh number, $\mathrm{Pr}=\nu/\chi$ is the Prandtl number, and $\mathbf{e}_z$ is the unit vector of the $z$-axis. For an incompressible two-dimensional flow~(\ref{eq:tc02}), one can introduce the stream function $\psi$:
\[
v^{(x)}=-\frac{\partial\psi}{\partial z}\,,
\qquad
v^{(z)}=\frac{\partial\psi}{\partial x}\,.
\]
Hence, the difference between the partial $x$-derivative of the $z$-component of (\ref{eq:tc04}) and the partial $z$-derivative of its $x$-component yields
\begin{equation}
\triangle\frac{\partial\psi}{\partial t}
=\triangle^2\psi+\mathrm{Ra}\frac{\partial\theta}{\partial x}\,,
\label{eq:tc06}
\end{equation}
and equation~(\ref{eq:tc05}) takes form
\begin{equation}
\frac{\partial\theta}{\partial t}
=\frac{1}{\mathrm{Pr}}\triangle\theta+\frac{s}{\mathrm{Pr}}\frac{\partial\psi}{\partial x}\,.
\label{eq:tc07}
\end{equation}

We consider the case of fixed temperature boundary conditions~\cite{Rayleigh-1916}:
\[
\theta|_{z=0}=\theta|_{z=1}=0\,.
\]
For simplicity of analytical calculations without the loss of fundamental generality of results, we consider inflexible shear-stress-free boundaries\footnote{In the case of the shear-stress-free boundary conditions, the special neutral mode is present: mean drift, motion with arbitrary constant velocity of the whole fluid in any horizontal direction. However, the existence of this mode does not affect the analysis presented in this paper.}, which yields
\[
\psi|_{z=0}=\left.\frac{\partial^2\psi}{\partial z^2}\right|_{z=0}=
\psi|_{z=1}=\left.\frac{\partial^2\psi}{\partial z^2}\right|_{z=1}=0\,.
\]
With adopted boundary conditions, equations~(\ref{eq:tc06})--(\ref{eq:tc07}) admit solutions of the form
\begin{equation}
\psi=\Psi(t)\cos{kx}\sin{\pi z}\,,
\quad
\theta=\Theta(t)\sin{kx}\sin{\pi z}\,.
\label{eq:Ps1Th1}
\end{equation}
The temporal evolution of ``amplitudes'' $\Psi(t)$ and $\Theta(t)$ is governed by the system of ordinary differential equations, which is valid also for a spatially uniform time-dependent gravity ac\-cel\-er\-a\-tion, $\mathrm{Ra}(t)=\mathrm{Ra}_0(1+\sigma\xi(t))$~\cite{Zenkovskaya-Simonenko-1966,Gershuni-Zhukhovitskii-Iurkov-1970,Gershuni-Lyubimov-1998}:
\begin{align}
\frac{\mathrm{d}}{\mathrm{d}t}\Psi
&=-\mathcal{D}\Psi-\frac{k\mathrm{Ra}_0}{\mathcal{D}}\Theta-\frac{k\sigma\mathrm{Ra}_0}{\mathcal{D}}\xi(t)\Theta\,,
\label{eq:tc08}
\\
\frac{\mathrm{d}}{\mathrm{d}t}\Theta
&=-\frac{sk}{\mathrm{Pr}}\Psi-\frac{\mathcal{D}}{\mathrm{Pr}}\Theta\,,
\label{eq:tc09}
\end{align}
where $\mathcal{D}\equiv\pi^2+k^2$, $\sigma$ is the noise strength. We consider a fast oscillating noise with short autocorrelation time, for which a natural mathematical idealization of $\delta$-correlated noise can be adopted: $\langle\xi\rangle=0$, $\langle\xi(t)\xi(t^\prime)\rangle=2\delta(t-t^\prime)$, where $\langle...\rangle$ indicates the averaging over noise realizations. According to the Central limit theorem, a $\delta$-correlated noise must be Gaussian (for other $\alpha$-stable noises $\int_{-\infty}^{+\infty}\mathrm{d}\tau\langle\xi(t)\xi(t+\tau)\rangle$ diverges). For stochastic equation system~(\ref{eq:tc08})--(\ref{eq:tc09}) we imply the Stratonovich interpretation.

It is also instructive to recall the stability properties of the noise-free ($\sigma=0$) system (\ref{eq:tc08})--(\ref{eq:tc09})~\cite{Rayleigh-1916}:
all perturbations decay for the heating from above, $s=-1$, and for the heating from below, $s=+1$, the perturbations with the wavenumber $k$ grow above $\mathrm{Ra}_\ast(k)=\mathcal{D}^3/k^2$. In an infinite layer, the instability threshold is
\[
\mathrm{Ra}_\mathrm{cr}=\frac{27\pi^4}{4}\,,
\]
where the perturbations with
\[
k_\mathrm{cr}(\sigma=0)=\frac{\pi}{\sqrt{2}}
\]
are marginally stable.

\section{Stochastic parametric excitation}\label{sec:spe}
Let us evoke the underlaying mathematics for the phenomenon of stochastic parametric excitation in linear systems~\cite{Klyatskin-2017,Klyatskin-1975,Khasminskii-2012}. For the linear stochastic differential equation system
\begin{equation}
\frac{\mathrm{d}}{\mathrm{d}t}x_j=\sum_{n=1}^N L_{jn}x_n+\sigma\xi(t)\sum_{n=1}^N G_{jn}x_n\,,
\;\; j=1,2,...,N,
\label{eq:spe01}
\end{equation}
written with the Stratonovich interpretation, one finds the Fokker--Planck equation governing the evolution of the probability density $w(\mathbf{x},t)$ averaged over noise realizations~\cite{Gardiner-1997}:
\begin{equation}
\frac{\partial w(\mathbf{x},t)}{\partial t}
 +\sum_j\frac{\partial}{\partial x_j}\bigg(\sum_n L_{jn}x_nw(\mathbf{x},t)\bigg)=\hat{Q}^2w(\mathbf{x},t)\,,
\label{eq:spe02}
\end{equation}
where $\hat{Q}(\cdot)\equiv\sum_j\frac{\partial}{\partial x_j}\left[\sum_n\sigma G_{jn}x_n(\cdot)\right]$. Multiplying equation~(\ref{eq:spe02}) by $x_l$ and integrating over all space $\mathbf{x}$, one finds
\begin{align}
\frac{\mathrm{d}}{\mathrm{d}t}\int\dots\int\mathrm{d}x_1\dots\mathrm{d}x_N\,x_lw
& +\sum_j\int\dots\int\mathrm{d}x_1\dots\mathrm{d}x_N\,x_l\frac{\partial}{\partial x_j}\bigg(\sum_n L_{jn}x_nw\bigg)
\nonumber
\\
& =\sum_j\int\dots\int\mathrm{d}x_1\dots\mathrm{d}x_N\,x_l\frac{\partial}{\partial x_j}\bigg(\sum_n \sigma G_{jn}x_n\hat{Q}w\bigg)\,.
\label{eq:spe03}
\end{align}
In the later equation for the average over noise realizations $\langle{x_l}\rangle\equiv\int\dots\int\mathrm{d}x_1\dots\mathrm{d}x_N\,x_lw$, with $w(\mathbf{x},t)$ decaying at infinity\footnote{For non-negative $w$, the integral over the sphere of radius $R$ $\oint_Rx_lw\mathrm{d}\mathcal{S}$ is of the same order of magnitude as the integral of $w$ over the space outside the sphere. The latter tends to $0$ for $R\to\infty$, as the integral of $w$ over all space is $1$.}, one can use partial integration to find
$$
\int\dots\int\mathrm{d}x_1\dots\mathrm{d}x_l\dots\mathrm{d}x_N\,x_l\frac{\partial}{\partial x_j}(...)=-\int\dots\int\mathrm{d}x_1\dots\mathrm{d}x_l\dots\mathrm{d}x_N\delta_{jl}(...).
$$
Performing partial integration once for the second term and twice for the last term of equation~(\ref{eq:spe03}), one obtains
\begin{equation}
\frac{\mathrm{d}}{\mathrm{d}t}\langle{x_j}\rangle =\sum_nL_{jn}\langle{x_n}\rangle +\sigma^2\sum_n(\mathbf{G}^2)_{jn}\langle{x_n}\rangle\,.
\label{eq:spe04}
\end{equation}

Matrix $\mathbf{G}^2$ can significantly change the eigenvalues of the problem and create the exponential growth of average fields in the systems where oscillations are conservative or decaying without noise.
Conceptually, this phenomenon was early on studied for a harmonic oscillator with stochastic modulation of the natural frequency $\ddot{x}+[\omega_0^2+\sigma\xi(t)]x=0$, where the noise was found to pump-in the oscillation energy~\cite{Klyatskin-2017,Klyatskin-1975}.

In literature~\cite{Klyatskin-2017,Klyatskin-1975}, this phenomenon is referred to as ``stochastic parametric resonance.'' However, we avoid using the word ``resonance'' here, since it suggests the matching of some fundamental time scales, as in the cases of (periodic modulation) parametric resonance~\cite{Faraday-1831,Rayleigh-1887,Landau-Lifshitz-2005}, stochastic resonance~\cite{Benzi-Sutera-Vulpiani-1981,Gang-etal-1993,Gammaitoni-etal-1998},
or coherence resonance~\cite{Pikovsky-Kurths-1997,Ushakov-etal-2005}, while here one observes no such matching. Technically, the effect we consider can be purely interpreted as the Stratonovich drift for energy-like functionals.

\section{Stochastic parametric excitations of thermal convection}\label{sec:spethc}
\begin{figure}[!t]
\center{
\includegraphics[width=0.62\textwidth]%
 {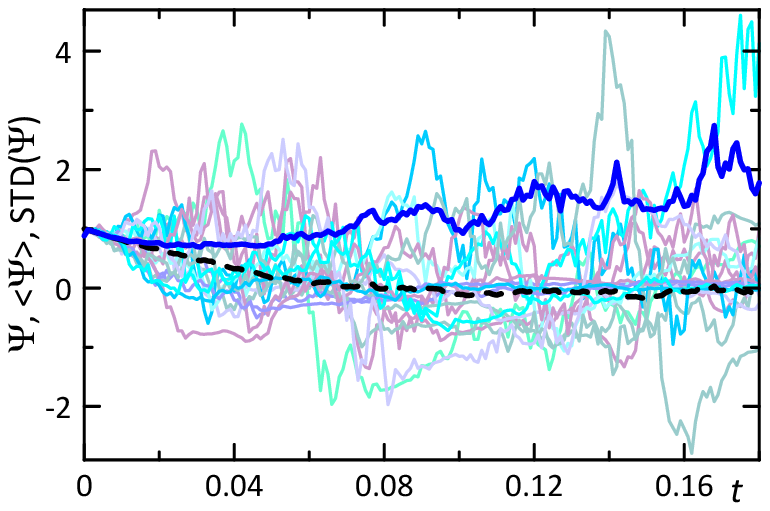}
}
\caption{The system dynamics above the stochastic excitation threshold is presented with $\Psi(t)$ for sample noise realizations (17 light-color curves), average $\langle\Psi(t)\rangle$ over 250 noise realizations (black dashed curve), and $\mathrm{STD}(\Psi)\equiv\sqrt{\langle\Psi^2(t)\rangle}$ for 250 noise realizations (bold blue solid curve). Parameters: $s=-1$, $k=\pi$, $\mathrm{Pr}=1$, $\mathrm{Ra}_0=\mathcal{D}^3/k^2=8\pi^4$, $\sigma^2=2\sigma_\ast^2=8/\mathcal{D}$.
Bold magenta line: the theoretical prediction $\mathrm{STD}(\Psi)\propto e^{\lambda_1t/2}$ with $\lambda_1\approx1.515\pi^2$ given by~(\ref{eq:stc11}); bold red line: the theoretical prediction $\langle\Psi(t)\rangle\propto e^{-2\pi^2t}\cos{2\pi^2t}$ given by (\ref{eq:stcm1})--(\ref{eq:stcm2}).
}
  \label{fig1}
\end{figure}

Comparing equation system~(\ref{eq:tc08})--(\ref{eq:tc09}) to (\ref{eq:spe01}), one can write down
\begin{equation}
\mathbf{L}=\left(
\begin{array}{cc}
  -\mathcal{D} & -\frac{k\mathrm{Ra}_0}{\mathcal{D}} \\[5pt]
  -\frac{sk}{\mathrm{Pr}} & -\frac{\mathcal{D}}{\mathrm{Pr}}
\end{array}\right),
\qquad
\mathbf{G}=\left(
\begin{array}{cc}
  0 & -\frac{k\mathrm{Ra}_0}{\mathcal{D}} \\[5pt]
  0 & 0
\end{array}\right).
\label{eq:stc01}
\end{equation}
Here one can calculate $\mathbf{G}^2=\mathbf{0}$. Therefore, the dynamics of average values given by~(\ref{eq:spe04}) is not affected by parametric noise:
\begin{align}
\frac{\mathrm{d}}{\mathrm{d}t}\langle\Psi\rangle
&=-\mathcal{D}\langle\Psi\rangle -\frac{k\mathrm{Ra}_0}{\mathcal{D}}\langle\Theta\rangle\,,
\label{eq:stcm1}
\\
\frac{\mathrm{d}}{\mathrm{d}t}\langle\Theta\rangle
&=-\frac{sk}{\mathrm{Pr}}\langle\Psi\rangle -\frac{\mathcal{D}}{\mathrm{Pr}}\langle\Theta\rangle\,.
\label{eq:stcm2}
\end{align}
However, this does not mean that noise is unable to induce convection. For instance, if noise induces oscillations with random shifts of the oscillation phase and these shifts tend to a uniform distribution, the superposition of solutions for many realizations can be small because of the symmetry. This superposition gives the noise-average, which can be small for nonsmall individual oscillation waveforms contributing to the superposition. In figure~\ref{fig1}, one can see how the superposition (the dashed line) of nondecaying noisy oscillations decays.

One should use here a more reliable statistical quantifier of the presence of convective motion. A reasonable choice for such quantifiers is $\langle\Psi^2\rangle$ and $\langle\Theta^2\rangle$, which cannot vanish because of the symmetry of the distribution for nondecaying convective motions. 
Early on, such an approach was demonstrated to be productive for describing the stochastic parametric excitation of a harmonic oscillator~\cite{Klyatskin-1975}, evaluating the free energy in the Ising model~\cite{Zillmer-Pikovsky-2005} and spatial localization exponents~\cite{Goldobin-Shklyaeva-2013} of convective patterns excited under frozen parametric disorder~\cite{Goldobin-Shklyaeva-2013,Goldobin-2019}.
From (\ref{eq:tc08})--(\ref{eq:tc09}) one can derive a self-contained equation system for $\Psi^2$, $\Psi\Theta$, and $\Theta^2$ of type (\ref{eq:spe01});
\begin{align}
\frac{\mathrm{d}}{\mathrm{d}t}\Psi^2
&
=-2\mathcal{D}\Psi^2-\frac{2k\mathrm{Ra}_0}{\mathcal{D}}\Psi\Theta -\frac{2k\sigma\mathrm{Ra}_0}{\mathcal{D}}\xi\Psi\Theta\,,
\label{eq:stc02}
\\
\frac{\mathrm{d}}{\mathrm{d}t}(\Psi\Theta)
&
=-\frac{sk}{\mathrm{Pr}}\Psi^2 -\frac{\mathrm{Pr}+1}{\mathrm{Pr}}\mathcal{D}\Psi\Theta
 -\frac{k\mathrm{Ra}_0}{\mathcal{D}}\Theta^2 -\frac{k\sigma\mathrm{Ra}_0}{\mathcal{D}}\xi\Theta^2,
\label{eq:stc03}
\\
\frac{\mathrm{d}}{\mathrm{d}t}\Theta^2
&
=-\frac{2sk}{\mathrm{Pr}}\Psi\Theta-\frac{2\mathcal{D}}{\mathrm{Pr}}\Theta^2.
\label{eq:stc04}
\end{align}
This equation system possesses form~(\ref{eq:tc01}) with  $\mathbf{x}=\{\Psi^2,\Psi\Theta,\Theta^2\}$ and
\begin{eqnarray}
&\mathbf{L}=\left(
\begin{array}{ccc}
  -2\mathcal{D} & -\frac{2k\mathrm{Ra}_0}{\mathcal{D}} & 0 \\[5pt]
  -\frac{sk}{\mathrm{Pr}} & -\frac{\mathrm{Pr}+1}{\mathrm{Pr}}\mathcal{D} & -\frac{k\mathrm{Ra}_0}{\mathcal{D}} \\[5pt]
  0 & -\frac{2sk}{\mathrm{Pr}} & -\frac{2\mathcal{D}}{\mathrm{Pr}}
\end{array}\right),
\label{eq:stc05}\\[5pt]
&\mathbf{G}=\left(
\begin{array}{ccc}
  0 & -\frac{2k\mathrm{Ra}_0}{\mathcal{D}} & 0 \\[5pt]
  0 & 0 & -\frac{k\mathrm{Ra}_0}{\mathcal{D}} \\[5pt]
  0 & 0 & 0
\end{array}\right),
\label{eq:stc06}\\[5pt]
&\mathbf{G}^2=\left(
\begin{array}{ccc}
  0 & 0 & \frac{2k^2\mathrm{Ra}_0^2}{\mathcal{D}^2} \\[5pt]
  0 & 0 & 0 \\[5pt]
  0 & 0 & 0
\end{array}\right).
\nonumber
\end{eqnarray}
Hence, employing equation~(\ref{eq:spe04}), one finds
\begin{align}
\frac{\mathrm{d}}{\mathrm{d}t}\langle\Psi^2\rangle
&
=-2\mathcal{D}\langle\Psi^2\rangle-\frac{2k\mathrm{Ra}_0}{\mathcal{D}}\langle\Psi\Theta\rangle +\frac{2\sigma^2k^2\mathrm{Ra}_0^2}{\mathcal{D}^2}\langle\Theta^2\rangle\,,
\label{eq:stc07}
\\
\frac{\mathrm{d}}{\mathrm{d}t}\langle\Psi\Theta\rangle
&
=-\frac{sk}{\mathrm{Pr}}\langle\Psi^2\rangle -\frac{\mathrm{Pr}+1}{\mathrm{Pr}}\mathcal{D}\langle\Psi\Theta\rangle -\frac{k\mathrm{Ra}_0}{\mathcal{D}}\langle\Theta^2\rangle\,,
\label{eq:stc08}
\\
\frac{\mathrm{d}}{\mathrm{d}t}\langle\Theta^2\rangle
&
=-\frac{2sk}{\mathrm{Pr}}\langle\Psi\Theta\rangle-\frac{2\mathcal{D}}{\mathrm{Pr}}\langle\Theta^2\rangle\,.
\label{eq:stc09}
\end{align}

The spectrum of exponential growth rates $\lambda_1\ge\lambda_2\ge\lambda_3$ of eigenvectors $\{\langle\Psi^2\rangle,\langle\Psi\Theta\rangle,\langle\Theta^2\rangle\}\propto e^{\lambda t}$ of the equations system~(\ref{eq:stc07})--(\ref{eq:stc09}) is given by the linear problem
\begin{equation}
\left(
\begin{array}{ccc}
 -2\mathcal{D}-\lambda & -\frac{2k\mathrm{Ra}_0}{\mathcal{D}} & \frac{2\sigma^2k^2\mathrm{Ra}_0^2}{\mathcal{D}^2} \\[5pt]
 -\frac{sk}{\mathrm{Pr}} & -\frac{\mathrm{Pr}+1}{\mathrm{Pr}}\mathcal{D}-\lambda & -\frac{k\mathrm{Ra}_0}{\mathcal{D}} \\[5pt]
 0 & -\frac{2sk}{\mathrm{Pr}} & -\frac{2\mathcal{D}}{\mathrm{Pr}}-\lambda
\end{array}
\right)
\left\{
\begin{array}{c}
 \langle\Psi^2\rangle \\[5pt]
 \langle\Psi\Theta\rangle \\[5pt]
 \langle\Theta^2\rangle
\end{array}
\right\}=0\,.
\label{eq:stc10}
\end{equation}
The characteristic equation for this linear problem reads
\begin{align}
\lambda^3 +3\mathcal{D}\frac{\mathrm{Pr}+1}{\mathrm{Pr}}\lambda^2 +\mathcal{D}^2\left(2+\frac{8}{\mathrm{Pr}}+\frac{2}{\mathrm{Pr}^2} -\frac{4sk^2\mathrm{Ra}_0}{\mathcal{D}^3\mathrm{Pr}}\right)\lambda
\qquad
\nonumber\\
+\frac{4(\mathrm{Pr}+1)}{\mathrm{Pr}^2}\left(\mathcal{D}^3-sk^2\mathrm{Ra}_0\right) -\frac{4\sigma^2k^4\mathrm{Ra}_0^2}{\mathcal{D}^2\mathrm{Pr}^2}=0\,.
\label{eq:stc11}
\end{align}

Now, let us note that equation system~(\ref{eq:stc02})--(\ref{eq:stc04}) is formally defined in an extended phase space $\{X,P,Y\}$ (where $X=\Psi^2$, $P=\Psi\Theta$, $Y=\Theta^2$) compared to the original physical space of $\Psi$ and $\Theta$; therefore, not all formal solutions of equation system~(\ref{eq:stc07})--(\ref{eq:stc09}) are physically meaningful. In particular, if one deals with an initial value problem, the initial state averaged over noise realizations is just the system state, i.e., the initial conditions must satisfy the relation $(\langle\Psi\Theta\rangle|_{t=0})^2=\langle\Psi^2\rangle|_{t=0}\langle\Theta^2\rangle|_{t=0}$. In the course of stochastic evolution the average values can deviate from the manifold given by this relation, but the solution still should obey some conditions. Specifically,
\begin{equation}
\langle\Psi^2\rangle\ge0\,,
\qquad
\langle\Theta^2\rangle\ge0\,.
\label{eq:stc12}
\end{equation}
There is also limitation on the value of $\langle\Psi\Theta\rangle$. As $(\alpha\Psi+\Theta)^2\ge0$ for all real $\alpha$, one can average $\alpha^2\Psi^2+2\alpha\Psi\Theta+\Theta^2\ge0$ and find $\alpha^2\langle\Psi^2\rangle+2\alpha\langle\Psi\Theta\rangle+\langle\Theta^2\rangle\ge0$. The latter inequality is fulfilled for all real $\alpha$ if the corresponding quadratic equation for $\alpha$ has no real roots or a pair of coinciding ones, i.e.,
\begin{equation}
\langle\Psi\Theta\rangle^2\le\langle\Psi^2\rangle\langle\Theta^2\rangle\,,
\label{eq:stc13}
\end{equation}
which is also known as the Cauchy--Schwarz inequality.
In particular, condition~(\ref{eq:stc12}) means that a physically meaningful solution of (\ref{eq:stc07})--(\ref{eq:stc09}) with the largest growth rate cannot be oscillatory. A physically observable instability for squares of fields should be strictly monotonous.

Cubic equation~(\ref{eq:stc11}) can be written in form $(\lambda-\lambda_1)(\lambda-\lambda_2)(\lambda-\lambda_3)$; therefore, the term with no $\lambda$ in polynomial~(\ref{eq:stc11}) equals $-\lambda_1\lambda_2\lambda_3$:
\[
\lambda_1\lambda_2\lambda_3=\frac{4\sigma^2k^4\mathrm{Ra}_0^2}{\mathcal{D}^2\mathrm{Pr}^2} -\frac{4(\mathrm{Pr}+1)}{\mathrm{Pr}^2}\left(\mathcal{D}^3-sk^2\mathrm{Ra}_0\right).
\]
Where the leading $\lambda_1$ is real-valued and passes from negative to positive values, either $\lambda_2$ and $\lambda_3$ are both negative or $\lambda_2=\lambda_3^\ast$; in either case, $\lambda_2\lambda_3>0$. Hence, the condition of existence of a monotonously growing mode is
\begin{equation}
\sigma^2>\sigma_\ast^2(k) =\frac{(\mathrm{Pr}+1)\mathcal{D}^2}{k^2\mathrm{Ra}_0^2}\left(\frac{\mathcal{D}^3}{k^2}-s\mathrm{Ra}_0\right).
\label{eq:stc14}
\end{equation}

A physically meaningful mode should obey conditions (\ref{eq:stc12}) and (\ref{eq:stc13}). From~(\ref{eq:stc10}) for $\lambda=0$, one can find
\[
\langle\Psi^2\rangle=\left(\frac{\sigma_\ast^2\mathrm{Ra}_0^2\mathrm{Pr}}{\mathrm{Pr}+1}\frac{k^2}{\mathcal{D}^3} +\frac{\mathcal{D}^2}{k^2}\right)\langle\Theta^2\rangle\,,
\]
which never violates (\ref{eq:stc12}), and $sk\langle\Psi\Theta\rangle=-\mathcal{D}\langle\Theta^2\rangle$, which yields
\[
\langle\Psi\Theta\rangle^2=\frac{\langle\Psi^2\rangle\langle\Theta^2\rangle} {1+ \frac{\sigma_\ast^2\mathrm{Ra}_0^2\mathrm{Pr}}{\mathrm{Pr}+1}\frac{k^4}{\mathcal{D}^5}}
\]
always satisfying condition~(\ref{eq:stc13}). Thus, the growing solution emerging at the threshold $\sigma_\ast$ is always physically meaningful.

In figure~\ref{fig1}, one can observe a decay of $\langle\Psi(t)\rangle$ averaged over nondecaying stochastic trajectories and see, that the value $\langle\Psi^2(t)\rangle$ is a good representative for the decay/growth of the intensity of stochastic oscillations.

\paragraph*{Initial value problem and the leading Eigenvector.}
Let us check, whether the noise-average system trajectory starting from the initial state $\{\Psi_0,\Theta_0\}$ must/can contain the leading formal Eigensolution of problem~(\ref{eq:stc10}), at least at the threshold of instability. For $\lambda=0$, the solution of the Hermitian conjugate of~(\ref{eq:stc10}) is $\{X^+,P^+,Y^+\}=\{\frac{k}{\mathrm{Pr}}, -2s\mathcal{D}, \frac{\mathrm{Pr}\mathcal{D}^2}{k}+\frac{\sigma_\ast^2k^3\mathrm{Ra}_0^2}{\mathcal{D}^3(\mathrm{Pr}+1)}\}$. For the initial system state $\{\Psi_0,\Theta_0\}$, the initial state of problem~(\ref{eq:stc02})--(\ref{eq:stc04}) is $\{\Psi_0^2,\Psi_0\Theta_0,\Theta_0^2\}$. The decomposition of this initial state into the basis of the Eigenvectors of problem~(\ref{eq:stc10}) does not contain the leading mode if
\[
\{X^+,P^+,Y^+\}\cdot\{\Psi_0^2,\Psi_0\Theta_0,\Theta_0^2\}=0\,.
\]
The latter equation can be recast as
\[
\left(\sqrt{\frac{k}{\mathrm{Pr}}}\Psi_0-s\mathcal{D}\sqrt{\frac{\mathrm{Pr}}{k}}\Theta_0\right)^2 +\frac{\sigma_\ast^2\mathrm{Ra}_0^2 k^3}{\mathcal{D}^3(\mathrm{Pr}+1)}\Theta_0^2=0\,,
\]
which admits only the trivial solution $\Psi_0=\Theta_0=0$. Thus, all initial states of the system contain a nonzero component of the leading formal mode; therefore, the growth of the formal leading mode of (\ref{eq:stc07})--(\ref{eq:stc09}) unambiguously dictates the growth of noisy physical system trajectories.

\begin{figure}[!t]
\center{
\includegraphics[width=0.56\textwidth]%
 {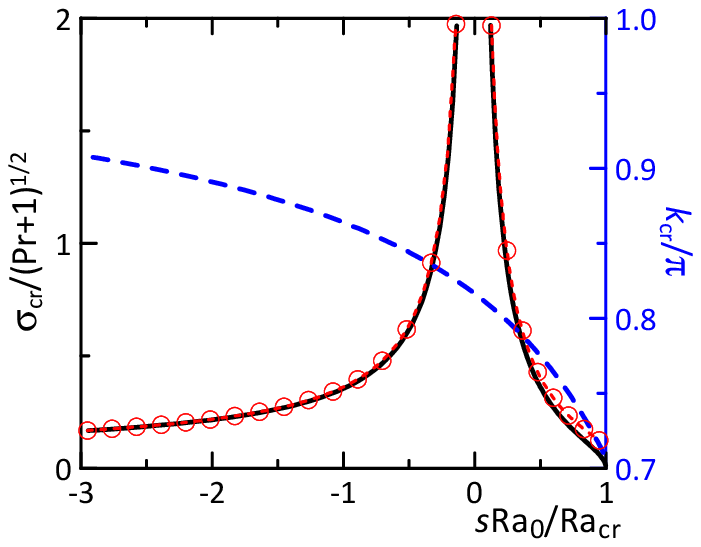}
}
\caption{The stochastic parametric excitation threshold $\sigma_\mathrm{cr}$ for an infinite layer is plotted with the black solid curves; the blue dashed curve shows the critical wavenumber $k_\mathrm{cr}$;
the asymptotic law~(\ref{eq:stc17}) 
is plotted with the red dotted curves marked by circles.}
  \label{fig2}
\end{figure}

\paragraph*{Stochastic excitation threshold.}
Minimizing $\sigma_\ast$ over $k$, one can find the critical values $k_\mathrm{cr}$ of the most easily excitable convective patterns and marginal noise strength $\sigma_\mathrm{cr}$ as functions of $\mathrm{Ra}_0$. Below the threshold of convective instability of the modulation-free system,
\[
s\mathrm{Ra}_0<\mathrm{Ra}_\mathrm{cr}\equiv\frac{27\pi^4}{4}\,,
\]
the dependence $\sigma_\ast(k)$ possesses a single local extremum. The dependence of the critical noise strength on $s\mathrm{Ra}_0$ can be provided in a parametric form:
\begin{align}
&\sigma_\mathrm{cr}^2=\frac{(\mathrm{Pr}+1)(2k_\mathrm{cr}^2-\pi^2)(\pi^2-k_\mathrm{cr}^2)} {(k_\mathrm{cr}^2+\pi^2)(2\pi^2-3k_\mathrm{cr}^2)^2}\,,
\label{eq:stc15}
\\
&s\mathrm{Ra}_0=\frac{(k_\mathrm{cr}^2+\pi^2)^3(2\pi^2-3k_\mathrm{cr}^2)} {k_\mathrm{cr}^2(\pi^2-k_\mathrm{cr}^2)}\,,
\label{eq:stc16}
\end{align}
where $\pi/\sqrt{2}\le k_\mathrm{cr}<\pi$. With these equations one can find an asymptotic law for $s\mathrm{Ra}_0\to-\infty$ (or $\pi-k_\mathrm{cr}\ll1$):
\begin{align}
\frac{\sigma_\mathrm{cr}^2}{\mathrm{Pr}+1}&\approx \frac{27\pi^6}{\mathrm{Ra}_0^2}\left(\frac{32}{27}-\frac{s\mathrm{Ra}_0}{\mathrm{Ra}_\mathrm{cr}}\right),
\label{eq:stc17}\\
k_\mathrm{cr}^2&\approx\pi^2\left(1-\frac{1}{\frac72-\frac{s\mathrm{Ra}_0}{\mathrm{Ra}_\mathrm{cr}}}\right).
\nonumber
\end{align}
The threshold (\ref{eq:stc15})--(\ref{eq:stc16}) and asymptotic law~(\ref{eq:stc17}) are presented in figure~\ref{fig2}. Noticeably, the asymptotic law is reasonably accurate not only for $s\mathrm{Ra}_0\to-\infty$, but for the entire parameter range where the noise-free system is convectively stable.

The stochastic excitation in the case of heating from above ($s=-1$) is not significantly less efficient than for heating from below ($s=+1$). Recall, the heating from above would make a purely stabilizing impact in the deterministic setups. Meanwhile, in our case, the bigger absolute value of the Rayleigh number results in a lower threshold of stochastic excitation, $\sigma_\mathrm{cr}^2\propto(\mathrm{Pr}+1)/\mathrm{Ra_0}$ (see equation~(\ref{eq:stc17})). For the low-gravity conditions the efficiency of the stochastic excitation requires subtle analysis; according to~(\ref{eq:stc17}),
$\sigma_\mathrm{cr}^2\propto1/\mathrm{Ra}_0^2$.

\paragraph*{Limit cases of strong heating from above and low-gravity conditions.}
In order to elucidate the limits of (i)~strong heating from above and (ii)~low-gravity, one must carry out the analysis in dimensional terms. First, we must link the dimensionless noise intensity $\sigma^2$ to the statistical properties of the fluctuating part $\widetilde{g}(t)\equiv g(t)-g_0$ of the gravity acceleration measured in experiments~\cite{Knabe-Eilers-1982}. Bearing in mind that $\sigma\xi(t)$ is a relative variation of the gravity acceleration and it is a white noise, we write
\[
2\sigma^2=\int\limits_{-\infty}^{+\infty}\langle\sigma\xi(t)\sigma\xi(t+t^\prime)\rangle\mathrm{d}t^\prime
=\int\limits_{-\infty}^{+\infty}\left\langle\frac{\widetilde{g}(t_\mathrm{dim})}{g_0} \frac{\widetilde{g}(t_\mathrm{dim}+t_\mathrm{dim}^\prime)}{g_0}\right\rangle\mathrm{d}\frac{t_\mathrm{dim}^\prime}{t_\ast}
=\frac{2\sigma_g^2}{g_0^2t_\ast}\,,
\]
where $t_\mathrm{dim}$ is a dimensional time, $t_\ast=h^2/\nu$ is the measure unit of the dimensionless time, and $\sigma_g^2$ is the intensity of gravity acceleration fluctuations in dimensional units. Hence,
\begin{equation}
\sigma^2=\frac{\nu\sigma_g^2}{g_0^2h^2}\,.
\label{eq:stc18}
\end{equation}
Second, we substitute the expressions for $\mathrm{Pr}$, $\mathrm{Ra}$, $\mathrm{Ra}_\mathrm{cr}$, $A=h\Delta T$ (where $\Delta T$ is the temperature difference between the horizontal boundaries) and equation~(\ref{eq:stc18}) into the asymptotic equations $\sigma_\mathrm{cr}^2\approx27\pi^6(\mathrm{Pr}+1)/(\mathrm{Ra}_0\mathrm{Ra}_\mathrm{cr})$ and $\sigma_\mathrm{cr}^2\approx32\pi^6(\mathrm{Pr}+1)/\mathrm{Ra}_0^2$. Hence, {\it for a strong heating from above},
\begin{equation}
\sigma_{g\,\mathrm{cr}}\approx2\pi\sqrt{\frac{(\nu+\chi)g_0}{\beta\Delta{T}h}}\,,
\label{eq:stc19}
\end{equation}
and, {\it for low-gravity conditions},
\begin{equation}
\sigma_{g\,\mathrm{cr}}\approx\frac{4\pi^3\sqrt{2\nu\chi(\nu+\chi)}}{\beta\Delta{T}h^2}\,.
\label{eq:stc20}
\end{equation}

According to (\ref{eq:stc19}), under a strong heating from above, the critical noise strength is lower for a stronger density heterogeneity across the layer $\beta\Delta T$ and a thicker layer. The dissipative resistance to excitation due to viscosity and heat diffusivity is represented by the combined measure $(\nu+\chi)$. The basic gravity acceleration $g_0$ increases the instability threshold, but this increase is slower than linear, $\sigma_{g\,\mathrm{cr}}\propto\sqrt{g_0}$\,.

According to (\ref{eq:stc20}), under the low-gravity conditions, the constant component of the residual gravity acceleration $g_0$ does not influence the leading order. The critical noise strength is lower for a stronger density heterogeneity $\beta\Delta T$ and a thicker layer. The dissipative resistance to excitation due to viscosity and heat diffusivity is represented by the combined measure $\nu\chi(\nu+\chi)$. The structure of the later combined measure is worthy of noticing. It becomes small, resulting in a low stochastic instability threshold, if either one of the two dissipation mechanisms, viscosity or heat diffusivity, becomes weak.


\section{Heat transfer: Nusselt number}
\label{sec:Nu}
The dimensionless heat transfer through the layer boundaries is characterized by the Nusselt number which is known to scale quadratically with the current amplitude moderately above the instability threshold~\cite{Gershuni-Zhukhovitskii-1976,Silveston-1958a,Silveston-1958b}. Therefore, the heat transfer across the layer can be characterized only within the framework of a nonlinear problem. Equation system~(\ref{eq:tc01})--(\ref{eq:tc03}) without linearization yields a nonlinear version of equations~(\ref{eq:tc06})--(\ref{eq:tc07}):
\begin{align}
\frac{\partial}{\partial t}\triangle\psi +\frac{1}{\mathrm{Pr}}J(\psi,\triangle\psi)
&=\triangle^2\psi+\mathrm{Ra}\frac{\partial\theta}{\partial x}\,,
\label{eq:Nu01}
\\
\frac{\partial\theta}{\partial t} +\frac{1}{\mathrm{Pr}}J(\psi,\theta)
&=\frac{1}{\mathrm{Pr}}\triangle\theta+\frac{s}{\mathrm{Pr}}\frac{\partial\psi}{\partial x}\,,
\label{eq:Nu02}
\end{align}
where Jacobian
\[
J(f,g)\equiv\frac{\partial f}{\partial x}\frac{\partial g}{\partial z}
 -\frac{\partial f}{\partial z}\frac{\partial g}{\partial x}\,.
\]

At an impermeable boundary, the normal component of the current velocity $v^{(z)}=0$ and the dimensionless heat flux
$J_Q=-\mathrm{Pr}^{-1}\frac{\partial}{\partial z}(-sz+\theta)\big|_{z=0}=s\mathrm{Pr}^{-1}(1-s\frac{\partial}{\partial z}\theta)|_{z=0}$. Hence, the Nusselt number
\begin{equation}
\mathrm{Nu}=1-s\left\langle\left.\frac{\partial\theta}{\partial z}\right|_{z=0}\right\rangle\,.
\label{eq:Nu_def}
\end{equation}
Equation~(\ref{eq:Ps1Th1}) gives the shape of the leading order linear solution
\[
\psi_1=\Psi(t)\cos{kx}\sin{\pi z}\,,
\quad
\theta_1=\Theta(t)\sin{kx}\sin{\pi z}\,,
\]
the average of which was found to decay to zero in the parameter range of our interest; therefore, the leading nonlinear correction $\{\psi_2,\theta_2\}$ is needed. Moreover, if $\mathrm{Nu}$ (\ref{eq:Nu_def}) is averaged over the boundary area, $\theta_1\propto\sin{kx}$ also makes a zero contribution.

For $\{\psi_2,\theta_2\}$, equation system~(\ref{eq:Nu01})--(\ref{eq:Nu02}) yields
\begin{align}
\frac{\partial}{\partial t}\triangle\psi_2 +\frac{1}{\mathrm{Pr}}J(\psi_1,\triangle\psi_1)
&=\triangle^2\psi_2+\mathrm{Ra}\frac{\partial\theta_2}{\partial x}\,,
\label{eq:Nu03}
\\
\frac{\partial\theta_2}{\partial t} +\frac{1}{\mathrm{Pr}}J(\psi_1,\theta_1)
&=\frac{1}{\mathrm{Pr}}\triangle\theta_2+\frac{s}{\mathrm{Pr}}\frac{\partial\psi_2}{\partial x}\,,
\label{eq:Nu04}
\end{align}
where $J(\psi_1,\triangle\psi_1)=0$ and
\[
J(\psi_1,\theta_1)=-\frac{k\pi}{2}\Psi(t)\,\Theta(t)\sin{2\pi z}\,.
\]
Hence,
\begin{equation}
\{\psi_2,\theta_2\}=\{\Psi_2(t),\Theta_2(t)\}\sin{2\pi z}
\label{eq:Ps2Th2}
\end{equation}
and equations~(\ref{eq:Nu03}) and (\ref{eq:Nu04}) become not only mutually decoupled but also the term with $\xi(t)$ (in $\mathrm{Ra}$) drops out:
\begin{align}
\frac{\mathrm{d}\Psi_2}{\mathrm{d}t}&=-4\pi^2\Psi_2\,,
\label{eq:Nu05}
\\
\frac{\mathrm{d}\Theta_2}{\mathrm{d}t}
&=-\frac{4\pi^2}{\mathrm{Pr}}\Theta_2+\frac{k\pi}{2\mathrm{Pr}}\Psi\Theta\,.
\label{eq:Nu06}
\end{align}
According to (\ref{eq:Nu05}), the flow $\Psi_2$ always decays; equation~(\ref{eq:Nu06}) can be averaged over noise realizations:
\begin{equation}
\frac{\mathrm{d}}{\mathrm{d}t}\langle\Theta_2\rangle
=-\frac{4\pi^2}{\mathrm{Pr}}\langle\Theta_2\rangle +\frac{k\pi}{2\mathrm{Pr}}\langle\Psi\Theta\rangle\,.
\label{eq:Nu07}
\end{equation}
For $\langle\Psi\Theta\rangle\propto e^{\lambda t}$, we consider $\langle\Theta_2\rangle\propto e^{\lambda t}$ and
\[
\langle\Theta_2\rangle =\frac{k\pi}{2}
\frac{\langle\Psi\Theta\rangle}{\lambda\mathrm{Pr}+4\pi^2}
=-\frac{\pi}{4s}
\frac{\lambda\mathrm{Pr}+2\mathcal{D}}{\lambda\mathrm{Pr}+4\pi^2} \langle\Theta^2\rangle
\,,
\]
where we used equation~(\ref{eq:stc09}).

Substituting temperature field from~(\ref{eq:Ps2Th2}) into expression~(\ref{eq:Nu_def}), we find
\begin{align}
\mathrm{Nu}
&=1-2\pi s\langle\Theta_2\rangle
\nonumber\\
&=1+\frac{\pi^2}{2}
\frac{\lambda\mathrm{Pr}+2(\pi^2+k^2)}{\lambda\mathrm{Pr}+4\pi^2}\langle\Theta^2\rangle
\label{eq:Nu08}\\
&=1+\frac{\pi^2}{2}
\frac{\lambda\mathrm{Pr}+2(\pi^2+k^2)}{\lambda\mathrm{Pr}+4\pi^2}\big[\Theta(0)\big]^2e^{\lambda t}.
\label{eq:Nu09}
\end{align}
Here we do not consider the nonlinear saturation of the perturbation growth. For a statistically stationary state, where nonlinear terms prevent the further exponential growth of convective currents, it is not sufficient to merely set $\lambda=0$ in equation~(\ref{eq:Nu08}) to reflect the fact that $(\mathrm{d}/\mathrm{d}t)\langle\Theta^2\rangle=0$ in (\ref{eq:stc09}) and $(\mathrm{d}/\mathrm{d}t)\langle\Theta_2\rangle=0$ in~(\ref{eq:Nu07}), but a more subtle consideration is required.

Thus, the exponential growth rate of $\langle\Theta^2\rangle$ we explored in previous sections determines the excitation of the convective transfer of heat which is an essential observable for thermal convection. This observable also plays a crucial role in complex systems incorporating the heat transfering/insulating elements.

\section{Comparison to high-frequency periodic modulation}
\label{sec:periodic}
Consider the case of high-frequency modulation of parameters, where $\sigma\xi(t)=\sigma\Omega\cos{\Omega t}$ in equation system~(\ref{eq:spe01}), $\Omega\gg|L_{jk}|$~\cite{Zenkovskaya-Simonenko-1966,Gershuni-Zhukhovitskii-Iurkov-1970,Gershuni-Lyubimov-1998}.
Not only the deterministic vibrational fluid dynamics is a versatile and well developed subject area~\cite{Zenkovskaya-Simonenko-1966,Gershuni-Zhukhovitskii-Iurkov-1970,Gershuni-Lyubimov-1998, Schlichting-Gersten-2000,Wolf-1969,Lyubimov-Cherepanov-1986,Shevtsova-etal-2016, Benjamin-Ursell-1954,Kumar-Tuckerman-1994,Shklyaev-Khenner-Alabuzhev-2008,Shklyaev-Alabuzhev-Khenner-2009, Nepomnyashchy-Simanovskii-2013,Goldobin-etal-2014,Goldobin-etal-2015,Pimenova-etal-2018, Klimenko-Lyubimov-2012,Klimenko-2017,Klimenko-2018,Omoteso-etal-2021}, but also in the field of nonlinear dynamics the phenomenon of stochastic resonance found its deterministic counterpart---the vibrational resonance~\cite{Landa-McClintock-2000,vibrstochres-2021a,vibrstochres-2021b},
which was also studied in fluid dynamical systems~\cite{Omoteso-etal-2021}.

\paragraph{Derivation of average equations for high-frequency periodic modulation.}
With the standard multiple scale method~\cite{Nayfeh} and $\varepsilon=\Omega^{-1}$, $\frac{\mathrm{d}}{\mathrm{d}t}=\varepsilon^{-1}\frac{\partial}{\partial t_{-1}}+\frac{\partial}{\partial t_0}+\dots$, $\mathbf{x}=\mathbf{x}^{(0)}+\varepsilon\mathbf{x}^{(1)}+\dots$, one can write
\begin{align}
&\left(\varepsilon^{-1}\frac{\partial}{\partial t_{-1}}+\frac{\partial}{\partial t_0}+\dots\right)(\mathbf{x}^{(0)}+\varepsilon\mathbf{x}^{(1)}+\dots)
\nonumber\\
&\qquad
{}=\mathbf{L}\cdot(\mathbf{x}^{(0)}+\varepsilon\mathbf{x}^{(1)}+\dots)
+\varepsilon^{-1}\sigma_0\cos{t_{-1}}\mathbf{G}\cdot(\mathbf{x}^{(0)}+\varepsilon\mathbf{x}^{(1)}+\dots)\,.
\label{eq:hfm01}
\end{align}

In the order $\varepsilon^{-1}$:
\begin{equation}
\frac{\partial\mathbf{x}^{(0)}}{\partial t_{-1}}=\sigma_0\cos{t_{-1}}\mathbf{G}\cdot\mathbf{x}^{(0)}.
\label{eq:hfm02}
\end{equation}
Hence,
\[
\mathbf{x}^{(0)}=e^{q\mathbf{G}}\cdot\mathbf{y}(t_0,t_1,\dots)\,,
\qquad
q=\sigma_0\sin{t_{-1}}\,,
\]
where $\exp\mathbf{A}\equiv\mathbf{I}+\mathbf{A}+\mathbf{A}^2/2!+\mathbf{A}^3/3!+\dots$, $\mathbf{I}$ is the identity matrix, and $\mathbf{y}(t_0,t_1,\dots)$ is an arbitrary vector constant in fast time $t_{-1}$.

Hermitian conjugate problem for (\ref{eq:hfm02}) follows from
$\int_0^{2\pi}\mathbf{z}\cdot(\frac{\partial}{\partial t_{-1}}\mathbf{x}^{(0)}-\sigma_0\cos{t_{-1}}\mathbf{G}\cdot\mathbf{x}^{(0)})\mathrm{d}t_{-1}=0$,
which can be rewritten as
$\int_0^{2\pi}(-\frac{\partial}{\partial t_{-1}}\mathbf{z}-\sigma_0\cos{t_{-1}}\mathbf{z}\cdot\mathbf{G})\cdot\mathbf{x}^{(0)}\mathrm{d}t_{-1}=0$ and yields
\begin{equation}
\mathbf{z}=\mathbf{z}_0\cdot e^{-q\mathbf{G}}\,,
\label{eq:hfm03}
\end{equation}
where $\mathbf{z}_0$ is an arbitrary constant vector.

In the order $\varepsilon^0$:
\[
\frac{\partial\mathbf{x}^{(1)}}{\partial t_{-1}} +\frac{\partial\mathbf{x}^{(0)}}{\partial t_0}
=\mathbf{L}\cdot\mathbf{x}^{(0)}
+\sigma_0\cos{t_{-1}}\mathbf{G}\cdot\mathbf{x}^{(1)}.
\]
Multiplying the latter equation by (\ref{eq:hfm03}) and averaging over $t_{-1}$, one finds
\[
\frac{\partial}{\partial t_0}\overline{\mathbf{z}_0\cdot e^{-q\mathbf{G}}\cdot\mathbf{x}^{(0)}}
=\overline{\mathbf{z}_0\cdot e^{-q\mathbf{G}}\cdot\mathbf{L}\cdot\mathbf{x}^{(0)}},
\]
where $\overline{(\dots)}$ indicates averaging over fast time $t_{-1}$. Omitting an arbitrary vector $\mathbf{z}_0$ and substituting $\mathbf{x}^{(0)}$, one obtains
\begin{equation}
\frac{\partial}{\partial t_0}\mathbf{y}
=\overline{e^{-q\mathbf{G}}\cdot\mathbf{L}\cdot e^{q\mathbf{G}}}\cdot\mathbf{y}\,.
\label{eq:hfm04}
\end{equation}

For specific problems (e.g., \cite{Landa-McClintock-2000,Omoteso-etal-2021}), matrix $\mathbf{G}$ is often sparse and possesses only zero eigenvalues. With all eigenvalues equal to zero, matrix $\mathbf{G}^n=0$ for large enough $n$ (e.g., in~(\ref{eq:stc01}) $\mathbf{G}^2=0$, in~(\ref{eq:stc06}) $\mathbf{G}^3=0$); therefore, the series representation of $\exp(q\mathbf{G})$ contains a finite number of terms. For $\mathbf{G}^3=0$ {\em or} keeping only $\sigma^2$-terms in (\ref{eq:hfm04}), one can calculate the average $\overline{e^{-q\mathbf{G}}\cdot\mathbf{L}\cdot e^{q\mathbf{G}}}$ in (\ref{eq:hfm04}) and find
\begin{equation}
\frac{\mathrm{d}\mathbf{y}}{\mathrm{d}t}
=\left[\mathbf{L} -\frac{\sigma^2}{2}\left(\mathbf{G}\cdot\mathbf{L}\cdot\mathbf{G} -\frac{\mathbf{L}\cdot\mathbf{G}^2+\mathbf{G}^2\cdot\mathbf{L}}{2}\right)\right]\cdot\mathbf{y}\,.
\label{eq:hfm05}
\end{equation}
Observed solution $\mathbf{x}(t)$ smoothed over fast oscillations
\begin{equation}
\overline{\mathbf{x}(t)} 
=\overline{e^{q\mathbf{G}}}\cdot\mathbf{y} =\left(\mathbf{I}+\frac{\sigma^2}{4}\mathbf{G}^2+\dots\right)\cdot\mathbf{y}(t)\,.
\label{eq:hfm06}
\end{equation}

Notice the difference between equation (\ref{eq:spe04}) for a stochastic modulation and equations (\ref{eq:hfm04}) (or (\ref{eq:hfm05})) and (\ref{eq:hfm06}) for a high-frequency periodic modulation.

\paragraph{Convection under vertical periodic vibrations.} For the problem~(\ref{eq:tc08})--(\ref{eq:tc09}) with matrices~(\ref{eq:stc01}), one finds
\[
\mathbf{L} -\frac{\sigma^2}{2}\mathbf{G}\cdot\mathbf{L}\cdot\mathbf{G}
=\left(
\begin{array}{cc}
  -\mathcal{D} & -\frac{k\mathrm{Ra}_0}{\mathcal{D}} +\frac{sk^3\sigma^2\mathrm{Ra}_0^2}{2\mathcal{D}^2\mathrm{Pr}}\\[5pt]
  -\frac{sk}{\mathrm{Pr}} & -\frac{\mathcal{D}}{\mathrm{Pr}}
\end{array}\right)
\]
and $\overline{e^{q\mathbf{G}}}\cdot\mathbf{y}=\mathbf{y}$. The characteristic equation for the exponential growth rate $\lambda$ reads
\[
\lambda^2+\mathcal{D}\frac{\mathrm{Pr}+1}{\mathrm{Pr}}\lambda
+\frac{\mathcal{D}^2}{\mathrm{Pr}} +\frac{k^4\sigma^2\mathrm{Ra}_0^2}{2\mathcal{D}^2\mathrm{Pr}^2} -\frac{sk^2\mathrm{Ra}_0}{\mathcal{D}\mathrm{Pr}}=0\,.
\]
Hence, the critical Rayleigh number for perturbations of wavenumber $k$ is lifted by vibrations:
\[
s\mathrm{Ra}_\mathrm{cr}=s\mathrm{Ra}_\mathrm{cr,\sigma=0} +\frac{\sigma^2k^2\mathrm{Ra}_\mathrm{cr}^2}{2\mathcal{D}\mathrm{Pr}}\,.
\]
Periodic vibrations make a purely stabilizing effect on thermal convection in this system~\cite{Zenkovskaya-Simonenko-1966,Gershuni-Zhukhovitskii-Iurkov-1970,Gershuni-Lyubimov-1998}.

\paragraph{Comparison of efficiencies of stochastic and high-frequency periodic modulations.}
It is also instructive to discuss the difference between equation system~(\ref{eq:spe04}) for the stochastic modulation and equation system~(\ref{eq:hfm05})--(\ref{eq:hfm06}) for the high-frequency periodic modulation. The latter is obviously much more sophisticated.
An essential dissimilarity between the action mechanisms of the stochastic and high-frequency periodic modulations makes the task of their comparison nontrivial and somewhat voluntaristic. Below we consider two approaches to this task.

At first glance, the shape of equation system~(\ref{eq:spe04}) can be expected also for the case of periodic vibrations. If one takes $\sigma\xi(t)=\sigma_1\cos\Omega t$, the perturbation expansion with the small parameter $\Omega^{-1}$ yields
\begin{equation}
\frac{\mathrm{d}}{\mathrm{d}t}\overline{\mathbf{x}(t)}
=\left[\mathbf{L} +\frac{\sigma_1^2}{\Omega}
\overline{\cos\Omega t(-\sin\Omega t)}
\mathbf{G}^2+\dots\right]\cdot\overline{\mathbf{x}(t)} \,,
\label{eq:hfm07}
\end{equation}
where ``$\dots$'' stand for higher order terms; but the $\sigma_1^2$-term vanishes as the average $\overline{\cos\Omega t\sin\Omega t}=0$. If the $\sigma_1^2$-term was not canceled, equation system~(\ref{eq:hfm07}) with $\sigma_1=\sigma\sqrt{\Omega}$ would have the same form as (\ref{eq:spe04}), with the same drift nature of the $\sigma^2$-term. The term is cancelled and we should take a stronger modulation, $\sigma\Omega\cos\Omega t$ (instead of $\sigma\sqrt{\Omega}\cos\Omega t$), and arrive at equation system~(\ref{eq:hfm05})--(\ref{eq:hfm06}) in a more sophisticated way.
The necessity to take a stronger periodic modulation
might suggest that the high-frequency periodic modulation is a weaker mechanism, as compared to the stochastic one.

Alternatively, specifically for mechanical systems, one can employ an energetic approach. This approach is also not fully straightforward, as the white noise forcing pumps the mechanical energy into the system while the high-frequency periodic does zero average work. Therefore, we cannot compare these mechanisms equitably; instead we can quantify the intensity of the stochastic action by the mechanical energy gain per one dimensionless unit of time and the intensity of the periodic action by the kinetic energy of the vibrational motion of the system. For the convective system we consider, the rate of kinetic energy-density gain from the stochastic modulation is
\begin{align}
\left\langle\frac{\mathrm{d}}{\mathrm{d}t}\frac{\rho_0u^2}{2}\right\rangle &=\rho_0\left\langle{u\frac{\mathrm{d}u}{\mathrm{d}t}}\right\rangle =\rho_0\Big\langle{\big(g(t)-g_0\big)\int_{-\infty}^t\mathrm{d}t_1\big(g(t_1)-g_0\big)}\Big\rangle
\nonumber\\
&=\rho_0g_0^2\sigma^2\Big\langle\xi(t)\int_{-\infty}^t\mathrm{d}t_1\xi(t_1)\Big\rangle =\rho_0g_0^2\sigma^2,
\label{eq:hfm08}
\end{align}
where $u$ is the layer/container velocity; under the periodic modulation $\sigma\xi(t)=\sigma\Omega\cos\Omega t$ the kinetic energy density is
\begin{align}
\overline{\frac{\rho_0u^2}{2}} &=\frac{\rho_0}{2}\overline{\left(\int^t\mathrm{d}t_1\big(g(t_1)-g_0\big)\right)^2}
=\frac{\rho_0g_0^2\sigma^2}{2}\overline{\left(\int^t\mathrm{d}t_1\xi(t_1)\right)^2} =\frac{\rho_0g_0^2\sigma^2}{4}\,.
\label{eq:hfm09}
\end{align}
Hence, energetically, the normalization $\xi(t)=\Omega\cos\Omega t$ is consistent with the normalization of the white noise (up to a normalization factor of $4$).

The energetic approach can be generalized for equation system~(\ref{eq:spe01}); one can introduce an energy-like function of state
\[
E\equiv\sum_{j=1}^N\frac{x_j^2}{2}
\]
and obtain the $E$-gain from the white noise modulation
\[
\left\langle\frac{\mathrm{d}E}{\mathrm{d}t}\right\rangle =\sigma^2\big\langle|\mathbf{G}\cdot\mathbf{x}|^2\big\rangle
\]
and the vibrational part of $E$ for periodic modulation $\xi(t)=\Omega\cos\Omega t$\,:
\[
\overline{\frac{|\mathbf{x}|^2}{2}-\frac{|\overline{\mathbf{x}}|^2}{2}} =\frac{\sigma^2}{4}|\mathbf{G}\cdot\mathbf{\overline{x}}|^2.
\]
To conclude, a generalized energetic approach suggests that the normalization $\xi(t)=\Omega\cos{\Omega t}$ employed for the treatment of the high-frequency periodic modulation case is consistent with the white noise normalization. The slipping-away ``would-be'' similarity of the average equations for the normalization $\xi(t)=\sqrt{\Omega}\cos\Omega t$ is an elusive argument for the choice of normalization. With a consistent normalization, the stochastic and high-frequency periodic modulations make impacts of the same order of magnitude and thus their efficiency as control instruments is commensurable.

\section{Conclusion}
\label{sec:concl}
The difference between the cases of stochastic and high-frequency periodic modulation is much more fundamental than just different equations (\ref{eq:spe04}) and (\ref{eq:hfm04}). For a periodic modulation, the actual solution follows the average one (averaging over short-time oscillations) in a regular way; the decay of average solutions indicates the decay of exact solutions. On the contrary, for a stochastic modulation, the decay of the noise-average solution does not indicate the decay of convection in the system; fluctuating solutions can still grow as in figure~\ref{fig1}.

Considering the phenomenon of stochastic parametric excitation of convective currents in hydrodynamic systems, one should clearly distinguish this problem from the problems of enforced fluctuations, like the thermal ones.
Mathematically, one can recall the synchronization by common noise~\cite{Pikovsky-1984b,Teramae-Tanaka-2004,Goldobin-Pikovsky-2005a,Nakao-etal-2005,Goldobin-etal-2010,Goldobin-Dolmatova-2019}, where the common noise induces mutual coherence between the instantaneous states of the oscillators driven by common noise, while an intrinsic (thermal) noise counteracts this coherence~\cite{Goldobin-Pikovsky-2005b,Nakao-etal-2007}. Similarly, with the Anderson localization~\cite{Anderson-1958,Maynard-2001}, the frozen parametric disorder leads to spatial localization of waves, while time-dependent perturbations of parameters destroy the localization.

The case of nearly conservative systems with small dissipation due to viscosity and no energy sources deserves special remark. The action of high-frequency periodic vibrations can be often represented by an average conservative mass force field~\cite{Gershuni-Lyubimov-1998}. Such a force, like the conservative force of gravity, can pump-in the mechanical energy into system via interaction with the density defects, which are simultaneously brought into the system and removed from it (e.g., temperature defects created by mixing currents against the background of the heat flux through the system). Without active manipulation with the density, high-frequency vibrations cannot become a source of the mechanical energy of the average motion. Hence, high-frequency periodic vibrations in a nearly conservative fluid system (e.g.\ single-component isothermal fluid) with a slow viscosity-related dissipation cannot excite growing currents. On the contrary, random parametric modulation breaks the time-shift symmetry (both continuous and discrete) making the energy conservation law impossible (Noether's theorem). As a result, the random modulation can excite the macroscopic motion where the high-frequency periodic vibrations would make a purely conservative impact.

We derived the conditions for the stochastic parametric excitation of convective heat transfer (see equations~(\ref{eq:stc15})--(\ref{eq:stc16}), an accurate approximation~(\ref{eq:stc17}), and figure~\ref{fig2}). The convective heat transfer quantified by the Nusselt number is related to the second moments of the stream function and temperature perturbations~(\ref{eq:Nu08}). The stochastic excitation in the case of heating from below is not significantly more efficient than for heating from above, which would make a purely stabilizing impact in the deterministic setups. For the case of heating from above, the bigger absolute value of the Rayleigh number results in a lower threshold of stochastic excitations; the dimensional form of the stochastic excitation threshold is given by (\ref{eq:stc19}). For the low-gravity conditions, the stochastic excitation is still efficient (see equation~(\ref{eq:stc20})). Noteworthy, the stochastic excitation becomes easier if either one of the two dissipation mechanisms, viscosity or heat diffusivity, becomes weak.

To summarize, stochastic modulation of parameters can induce the excitation of macroscopic convective currents, where high-frequency modulation of the same strength can only suppress the convective instability.
The classical analytically solvable Rayleigh problem~\cite{Rayleigh-1916} is a convenient example for illustration, but the phenomenon is expected to be general.

\dataccess{This article has no additional data.}

\aucontribute{DSG conceived and designed the study. EVP and DSG performed all the analytical derivations presented in the paper independently to ensure their correctness, as well as obtained the numerical results presented in figure~2. Both authors interpreted the mathematical results,
read and approved the manuscript.}

\competing{The authors declare that they have no competing interests.}

\funding{The authors acknowledge financial support by the Ministry of Science and Higher Education of the Russian Federation (theme no.\ 121112200078-7).}


\end{document}